# Emergent giant topological Hall effect in twisted Fe$_3$GeTe$_2$ metallic system

Hyuncheol Kim[1,2,3†], Kai-Xuan Zhang[1,2,3†*$], Yu-Hang Li[4†], Giung Park[1,2,3], Ran Cheng[5,6], Je-Geun Park[1,2,3*]

[1]Department of Physics and Astronomy, Seoul National University, Seoul 08826, South Korea

[2]Center for Quantum Materials, Department of Physics and Astronomy, Seoul National University, Seoul 08826, South Korea

[3]Institute of Applied Physics, Seoul National University, Seoul 08826, South Korea

[4]School of Physics, Nankai University, Tianjin 300071, China

[5]Department of Electrical and Computer Engineering, University of California, Riverside, CA 92521, USA

[6]Department of Physics and Astronomy, University of California, Riverside, CA 92521, USA

$Present address: Department of Physics, Washington University in St. Louis, St. Louis, Missouri 63130, USA

† These authors contributed equally to this work

* Corresponding authors: Kai-Xuan Zhang (email: kxzhang.research@gmail.com), and Je-Geun Park (email: jgpark10@snu.ac.kr)

**Abstract**

**The topological Hall effect, driven by the exchange interaction between conduction electrons and topological magnetic textures such as skyrmions, is a powerful probe for investigating the topological properties of magnetic materials. Typically, this phenomenon arises in systems with broken global inversion symmetry, where Dzyaloshinskii-Moriya interactions stabilize such textures. Here, we report the discovery of an emergent giant topological Hall effect in the twisted Fe$_3$GeTe$_2$ metallic system, which notably preserves the general global inversion symmetry. This effect manifests exclusively within a narrow window of "magic" twist angles ranging from**

**0.45º to 0.75º, while it is absent identically outside of that range, highlighting its unique and emergent nature. Micromagnetic simulations reveal that this topological Hall effect originates from a skyrmion lattice induced by alternating in-plane and layer-contrasting Dzyaloshinskii-Moriya interactions that result from local inversion symmetry breaking. Our findings underscore twisted Fe$_3$GeTe$_2$ as a versatile platform for engineering and controlling topological magnetic textures in metallic twisted van der Waals magnets, thereby opening up new avenues for next-generation spintronic devices.**

**Introduction**

In non-centrosymmetric magnetic materials with broken global inversion symmetry, the interplay among Heisenberg exchange interaction, symmetry-enabled Dzyaloshinskii-Moriya interaction (DMI), and the uniaxial magnetic anisotropy can stabilize chiral spin textures such as skyrmions[1]. These topological magnetic structures hold great potential for achieving high-density storage and low-power computing[2]. Skyrmions can provide an effective gauge field that exerts a Lorentz-like force on moving electrons, giving rise to a transverse Hall current- a phenomenon known as the topological Hall effect (THE)[3,4]. This effect is a key signature for identifying and investigating the topological magnetic textures.

On the other hand, in centrosymmetric magnetic materials preserving global inversion symmetry, skyrmions could also emerge through mechanisms such as magnetic frustration[5,6] the Ruderman-Kittel-Kasuya-Yosida (RKKY) interaction[7], etc., hence giving rise to the THE. This phenomenon has been observed across various materials, including ferromagnetic semiconductors[8,9], oxides[10-15], transition metals[16-20], magnetic topological insulators[12,21,22], magnetic heterostructure[23-25], etc. Additionally, the global inversion symmetry in such centrosymmetric systems does not preclude local inversion symmetry breaking that allows for alternating DMIs on the atomic scale, hence the formation of skyrmions and the ensuing THE[26,27]. Despite these theoretical predictions, direct experimental observation of such intrinsic electronic responses in centrosymmetric systems remains elusive.

This article reports a giant THE in a centrosymmetric $Fe_3GeTe_2$ (FGT) twisted lattice respecting a global inversion symmetry, observed almost without an external magnetic field. As a representative member of the developing van der Waals (vdW) magnets[28-35], FGT[36-39] is a ferromagnetic metal with magnetic iron atoms arranged on a triangular lattice, preserving the spatial inversion symmetry. Twisting FGT to form an FGT/FGT homostructure creates a twisted lattice exhibiting alternating AA and AB stacking regions. Our calculation by using the Levy and Fert's model shows that twisting FGT can induce very strong intralayer DMI. This twisted FGT system introduces an excellent platform for studying emergent magnetic orderings driven by the competition among the exchange interaction, DMI, RKKY interaction, Coulomb interaction, and others[40,41].

Moreover, the itinerant electrons in FGT allow for exploring the topological nature of the emergent non-coplanar magnetic ordering through transport measurement. Unlike magnetic insulators such as $CrI_3$[42-45] and $CrSBr$[46,47], however, the strong metallic bonds in FGT pose a significant challenge for fabricating twisted lattices using conventional tear-and-stack techniques. To overcome this difficulty, we developed a novel PCL-hBN tear-and-stack method[48-50], leveraging the extreme stickiness of PCL to pick up the specific region of target materials like FGT while avoiding adhesion to hBN. This technique successfully fabricated a twisted FGT system with a twist angle from 0° to 5° (see Methods).

When the twist angle falls within the range of 0.45° to 0.75°, we observe a giant THE that is entirely absent at other angles, signaling the emergence of skyrmions. This effect is more pronounced in thinner FGT twisted systems, highlighting the strong interfacial influence at play. Furthermore, our micromagnetic simulations confirm that THE arises from intralayer DMI induced by twisting, which stabilizes the FGT twisted system into a skyrmion lattice within the specific twist angle range of 0.45° to 0.75°.

Our simulations further indicate that different magnetic ground states, including ferromagnetic metal and magnetic stripe phases, can be stabilized depending on the twist angle and the interlayer thickness. The twist-angle-dependent THE demonstrated in this work provides a compelling example of how topological spin texture can be created and manipulated using vdW ferromagnets. This discovery holds significant potential for various spintronics applications, such as employing a twisted FGT system as a magnetic quantum simulator or in high-density data storage.

## Results
**Twisted device fabrications and emergent topological Hall effect**

In light of the strong adhesion between PCL and FGT and the weaker interaction between hBN and FGT, we developed a combined PCL-hBN method to fabricate twisted FGT devices through an optimized tear-and-stack process, as illustrated in Fig. 1a-d. The process begins with preparing the PCL/hBN stamp (Fig. 1a), which picked up the h-BN nanoflake from a $SiO_2$/Si substrate.

Next, the PCL/hBN stamp is aligned with a selected FGT nanoflake on a separate $SiO_2$/Si substrate, ensuring that the hBN covers only half of the FGT nanoflake. Gradually increasing the temperature of the substrate to 62 °C allows the PCL polymer fringe to cross the entire FGT nanoflake fully. Then, the PCL/hBN stamp is lifted after the temperature has decreased to 30 °C to tear the FGT nanoflake. After this process, half of the FGT nanoflake covered by hBN remains on the $SiO_2$ substrate due to the weak adhesion of FGT to hBN. In contrast, the uncovered half FGT adheres to the PCL stamp due to the strong adhesive force of PCL (Fig. 1b). This results in the half FGT nanoflake (top FGT layer) being attached to the PCL stamp adjacent to the hBN. In contrast, the last half of the FGT nanoflake (bottom FGT layer) remains on the original $SiO_2$/Si substrate, completing the tear process.

The next step involves finely rotating the substrate to introduce a twist angle between the top FGT layer on the stamp and the bottom FGT layer on the $SiO_2$/Si substrate. Finally, the top FGT layer is carefully aligned on the bottom FGT layer and is gradually attached to the bottom FGT layer to complete the stacking process. Then, the temperature is increased above the melting point of PCL to detach the top FGT layer from the PCL stamp. After successful detachment, the substrate with FGT homostructure is dipped into tetrahydrofuran for 10 minutes to remove any PCL residual. As a result, the top FGT layer is well stacked onto the bottom FGT layer (Fig. 1c). This innovative technique overcomes the limitations of the conventional PC tear-and-stack method for rigid vdW metals, offering a significant advancement for the 2D vdW materials community.

Finally, Hall electrodes are patterned onto the overlapping region to facilitate the investigation of the transverse Hall response in the twisted region (Fig. 1d), forming the central focus of this study. Figures 1e and 1f present the schematic of the twisted FGT homostructure and the typical optical image of fabricated devices, respectively. The white scale bar represents approximately 5 μm.

In general, the transverse Hall resistance $R_{xy}$ consists of three components $R_{xy} = r_0 B + r_s M + R_T$. The first term represents the normal Hall effect arising from the Lorentz force induced by the magnetic field. The second term corresponds to the anomalous Hall effect (AHE), originating from spin-orbit coupling, where $r_0$ and $r_s$ are their respective coefficients. The third term $R_T$ is the THE, which results from the Berry phase accumulation in real space and is a hallmark of topological magnetic textures. When itinerant electrons traverse a topologically nontrivial spin texture, their spins undergo adiabatic evolution, acquiring an additional Berry phase due to the local spin configuration[51,52]. This Berry phase acts as an effective magnetic flux, giving rise to an anomalous velocity component transverse to the direction of electron motion. Consequently, the scalar spin chirality—defined as $\chi_{ijk} = \vec{S}_i \cdot (\vec{S}_j \times \vec{S}_k)$—produces a fictitious magnetic field that deflects the itinerant electrons, resulting in the observable topological Hall response. Since the normal Hall effect is linearly proportional to the applied magnetic field and the AHE can be determined from the magnetization, $R_T$ can, therefore, be extracted from the experimental data.

FGT is a hard ferromagnet of perpendicular magnetic anisotropy (See Supplementary Note 4[36,37,39]). As shown in Fig. 1g, a single pristine FGT nanoflake exhibits only a typical AHE at low temperatures, characterized by a ferromagnetic hysteresis loop in the magnetic field-dependent transverse resistance ($R_{xy}$-H curve)[53,54]. In contrast, a twisted FGT device with a twist angle of 0.6° (Fig. 1h) strikingly displays two asymmetric humps in addition to the typical anomalous Hall loop. These $R_{xy}$ humps are phenomenologically identical to the THE observed in the topological spin texture of a single magnetic system. This emergent THE is exclusive to metallic magnets, as it relies on electronic conduction, which is detectable through transport measurements. In the following sections, we systematically investigate this emergent quantum effect experimentally and elucidate its underlying mechanism induced by twisting.

**Twist angle dependence of the emergent giant topological Hall effect.**

Next, we investigate the dependence of the exotic THE on the twist angle. Figure 2 presents the normalized $R_{xy}$ (scaled by its maximum) as a function of the magnetic field (*H*) for various twisted FGT devices with different twist angles. All samples were fabricated under identical conditions, with the twist angle being the only controlled variable among the devices. As shown in Fig. 2a, the sample with a zero-twist angle exhibits the typical anomalous Hall loop at 20 K, consistent with the behavior of a pristine FGT nanoflake. Similarly, samples with twist angles of 0.15° and 0.3° display no topological Hall signals. Remarkably, however, a distinct topological Hall hump emerges in devices with twist angles of 0.45°, 0.6°, and 0.75°. As the twist angle increases beyond this range, the topological Hall hump disappears, and the pristine anomalous Hall loop reappears. Figure 2b demonstrates the same behavior even more prominently at a temperature of 60 K, confirming that the observed THE is indeed induced by twisting.

Before going further, we would like to discuss several important physical quantities for the THE in our twisted FGT system. Firstly, the supercell lattice $L_m(\theta)$ is related to the lattice constant *a* and a twist angle $\theta$. It can be expressed in the following formula: $L_m(\theta) = \frac{a}{2\sin(\theta/2)}$. For $\theta = 0.45°, 0.6°,$ and $0.75°$, this yields a supercell lattice of approximately 50.8, 38.1, and 30.5 nm, respectively. Secondly, the topological Hall resistivity $\rho_{xy}^{THE}$ can be related to the carrier density and the effective filed generated by the skyrmion lattice. Within this framework, $\rho_{xy}^{THE}$ can be written as $\rho_{xy}^{THE} = R_0 P \Phi_0 N_{sk}$, where $R_0$ is the ordinary Hall coefficient, *P* is the spin polarization, $\Phi_0$ is the magnetic flux quantum, and $N_{sk}$ is the skyrmion density[24,55]. Although this relation is oversimplified for thin-film magnetic layers with interfacial DMI, it remains useful for order-of-magnitude estimates[24,56]. In our twisted FGT with $\theta = 0.45°$, the magnitude of $\rho_{xy}^{THE}$ at 60 K is 0.71 μΩ·cm, and the corresponding $R_0$ is estimated to be 1.51×10⁻⁹ m³/C. Using the reported spin polarization of FGT, *P* is 0.53 at 60 K[57]. Then skyrmion density $N_{sk}$ is expected to be 2.15×10¹⁵ m⁻². For a triangular skyrmion lattice, this density is related to the skyrmion lattice constant *D* via $\frac{1}{N_{sk}} = \frac{\sqrt{3}}{2}D^2$, from which we estimate *D* is about 23 nm. These characteristics emerging on top of the ground-state magnetic configuration suggest a specific microscopic mechanism behind THE, which will be discussed in our theoretical analysis and calculation section afterwards.

To provide a concise representation, we define THE peak ratio as (topological Hall hump) / (topological Hall hump + anomalous Hall magnitude) × 100 %. This ratio is plotted as a color map in Fig. 2c, showing its dependence on temperature and twist angle. The THE peak ratio reaches values as high as 50 %, indicating a giant THE, even comparable in magnitude to the colossal AHE[37,38,58] previously reported for FGT. The latter arises from the large Berry curvatures associated with its topological nodal-line band structures[38,39,59]. As highlighted in the inset, we extended our examination of the twist angle up to 5°, comparable to the twist angle range adopted in the previous study of vdW magnetic insulators[42-44].

Notably, the twist-angle-dependent THE persists across a wide temperature range from 20 to 70 K (Fig.2c). These results not only confirm the universal dependence of the emergent THE on the twist-angle dependence but also demonstrate its reproducibility across different temperatures and individual devices at small twist angles.

**Thickness dependence of topological Hall effect**

The observed dependence on the twist angle suggests that this extraordinary THE is closely tied to the interfacial effect created by twisting. This inspires us to examine the thickness dependence of such behavior to investigate further the interfacial nature of the underlying interactions driving this phenomenon. We successfully achieved twisted FGT devices as thin as ~6 nm using our current fabrication method. Figure 3d illustrates the normalized $R_{xy}$ (scaled by its maximum) as a function of an out-of-plane magnetic field for several twisted FGT devices with the same twist angle of 0.75° but varying thicknesses, all measured at 50 K. The data clearly show a pronounced THE in the ~6 nm-thick device, consistent with the results in Fig. 2. However, as the thickness increases, the topological Hall hump becomes weaker in the ~10 nm-thick device before vanishing entirely in a ~20 nm-thick sample.

This behavior aligns with the intuitive expectation that the interfacial twist effect dominates (diminishes) in thinner (thicker) homostructures and is substantially suppressed in thick samples. The thickness dependence observed in Fig. 3d supports this hypothesis, with the disappearance of THE in the thickest 20 nm device, where the Hall response reverts to the typical AHE seen in pristine single FGT nanoflakes. Figures 3a-c and 3e-f display the same thickness-dependent trend across different temperatures.

Considering that the out-of-plane magnetic interaction in FGT expands to ~5 nm[36], this suppression of THE with increasing thickness highlights the critical role of the interfacial effect at the twisting interface. These observations shed significant light on the physical mechanism underlying this intriguing twist-induced THE, which we will further explore in the next section.

**Theoretical analysis**

To theoretically explain the emergent THE in the twisted FGT lattice, we analyze its underlying physical mechanism from an energy perspective. Since the THE is observed over a continuous twist angle range rather than at the discretized "magic angles" (Fig.2c), electron-electron interactions that dominate when the kinetic energy quenches at "magic angles", can be excluded as the primary mechanism for the skyrmion formation. Moreover, the absence of THE in pristine FGT as shown in Fig. 1g rules out the possibility of the superimposition of two ferromagnet with different coercive field. As detailed in the Supplementary Information, this twisted FGT lattice allows both intralayer and interlayer DMIs, which can be obtained by an analytical derivation based on the Lévy–Fert model within the perturbation theory framework (Supplementary Note 1). Our calculation shows that the intralayer DMI is about four orders of magnitude larger than the interlayer one. Meanwhile, the magnitude of the intralayer DMI is almost a constant independent of position inside the twisted FGT. Figure 4a shows the magnitude of the interlayer DMI as a function of twisting angle $\theta$ for three different thicknesses. It turns out that this intralayer DMI is significantly enhanced by twisting. Mathematically, this intralayer DMI can be expressed as $\vec{D}_{ij} \cdot \vec{S}_i \times \vec{S}_j$ where $\vec{D}_{ij} = D_0(\theta)(-1)^l \hat{z} \times \hat{e}_{ij}$ with $l = 1$ or 2 referring to the top or bottom layers, $\hat{z}$ being the unit vector of the plane normal, and $\hat{e}_{ij}$ the lattice vector connecting site $i$ and $j$.

In the presence of the DMI, the twisted FGT lattice can be modelled using the continuum free energy (Supplementary Note 2):

$$H = \int d\boldsymbol{r}^2 \{\sum_l [Ja_0^2(\nabla \vec{\boldsymbol{m}}_l)^2 - K(m_l^z)^2] + J_t \vec{\boldsymbol{m}}_1 \cdot \vec{\boldsymbol{m}}_2 \\ + D_0(\theta)a_0 \sum_l (-1)^l [\hat{\boldsymbol{y}} \cdot (\vec{\boldsymbol{m}}_l \times \partial_x \vec{\boldsymbol{m}}_l) - \hat{\boldsymbol{x}} \cdot (\vec{\boldsymbol{m}}_l \times \partial_y \vec{\boldsymbol{m}}_l)]\} \quad (1)$$

where $Ja_0^2$ is the exchange stiffness with $a_0$ the lattice constant, $K$ is the uniaxial anisotropy, $J_t$ is the interlayer exchange interaction, and $D_0(\theta)$ is the intralayer DMI constant obtained using the Lévy and Fert's model (Supplementary Note 1), which flips signs between layers and depends on the twisting angle $\theta$. Given the ferromagnetic nature of FGT, we resort to micromagnetic simulations to determine the ground-state magnetic configuration. Specifically, we discretize the energy in Eq. (1) on a triangular superlattice and numerically minimize the total energy at finite temperatures (see Methods). Figure 4b presents the phase diagram of the twisted FGT lattice representing by the skyrmion numbers versus the twist angle $\theta$ and interlayer thickness $d$, with the spin configurations at some representative $\theta$ and $d$ plotted in Fig. 4c. Generally, we identify three different phases on the $\theta - d$ plane, e.g., the magnetic stripe phase at small twisting angles, the ferromagnetic (FM) phase at large twisting angles, and the skyrmion lattice (SKX) phase interpolating the two. Meanwhile, we also find the magnetic stripe can coexist with the skyrmion lattice at some region.

When $d = 6$nm, the system is a skyrmion lattice at $0.45 \leq \theta \leq 0.75$. In principle, the topological Hall resistance is proportional to the spatial topological charge of the magnetic configuration: $\rho_{xy}^T = PQ/en$, where $P$ is the electron polarization, $Q$ is the topological charge of the magnetic ground state, and $e$ and $n$ are the electron charge and density, respectively[9,16]. Moreover, the THE peaks are typically proportional to the skyrmion density inside the skyrmion lattice. This phase diagram aligns quantitatively with the color map of the THE peak ratio shown in Fig. 2c. Although the spins in opposite layers rotate oppositely due to the reversed DMI, the topological charges in different layers are identical[27]. Moreover, the high density of the skyrmion lattice provides evidence for the giant THE peaks in the experimental data. Our simulations thus confirm that the giant THE can be ascribed to the emergence of skyrmion lattice at specific twisting angles.

Notably, the DMI plays a crucial role in facilitating the formation of skyrmions and spin stripes at small twist angles. Since the DMI depends heavily on the interlayer coupling induced by twisting, increasing the thickness of the twisted FGT system overall reduces the DMI as shown in Fig. 4a, thereby suppressing THE. Consequently, the THE only appears at a small twisting angle region indicated by the SKX in the phase diagram (Fig. 4b). Furthermore, competitions among the exchange interactions, uniaxial anisotropy, and DMI underpin the formation of skyrmions. Thermal agitations at high temperatures effectively reduce the uniaxial anisotropy, potentially enhancing the THE and broadening the range for the skyrmion formation, as observed in Fig. 2c.

**Discussion**

In our study, the emergence of THE in twisted FGT and micromagnetic simulation provide insights into how local inversion-symmetry breaking modifies the magnetic ground state through twisting. On this basis, we clarify several important issues on the emergent giant THE in our twisted FGT devices. First, the temperature dependence of the THE in twisted FGT suggests that the formation of topological spin textures is governed by a competition of magnetic anisotropy, exchange interactions, and the DMI (See Supplementary Note 5). Second, we estimated the skyrmion size from the magnitude of THE in the twisted system[24,55,56], which is consistent with the theoretical calculations (See Supplementary Note 6). Moreover, we find that the electronic mean free path in FGT is much smaller than the skyrmion size, which further supports our interpretation that the THE originates from topological spin textures (See Supplementary Note 7). Furthermore, we considered the cases in our experiments, where the "two-channel AHE"[60-62] is difficult to be straightforwardly applicable in our twisted FGT devices (See Supplementary Note 8).

In conclusion, we have observed a twist-angle-dependent THE in a FGT twisted lattice that preserves a general global inversion symmetry. The local inversion symmetry breaking induces an intralayer DMI, driving the transition from a ferromagnetic ground state to a skyrmion state and eventually to a spin stripe state as the twist angle decreases. The THE emerges as the system enters the skyrmion state, a phenomenon further corroborated by the micromagnetic simulations. Our findings demonstrated the potential to create and manipulate skyrmions and other spin states by engineering twisted magnetic systems. This study presents the first report of a twisted metallic vdW magnet enabled by our advanced fabrication technique and the first observation of an emergent THE at small twist angles. These results surpass previous studies on rare cases of twisted vdW magnetic insulators, highlighting the promise of twisted vdW magnetic metals for novel, current-driven spintronics applications.

**Methods**

**FGT single crystal growth**

High-quality FGT single crystals were grown using the chemical vapour transport (CVT) method, following our previous work[63]. High-purity chemical elements were purchased from Thermo Fisher (Fe 99.9%, Ge 99.999%, and Te 99.999%). These elements were placed into a quartz tube with low water and oxygen concentrations (below 1 ppm); all processes were carried out inside a glove box. Iodine was used as the transport agent for the CVT method. After all the chemical powders were placed into the quartz tube, the tube was evacuated and sealed. The sealed tube was then placed in the middle of a two-zone furnace. The temperatures of the cold and hot zones of the two-zone furnace were raised to 700 and 750°C, respectively. After two weeks, high-quality FGT single crystals were synthesized.

**Twisted FGT homostructure fabrication**

The detailed descriptions of the fabricating method have been presented in the main text and Fig. 1. Here, we would like to elaborate on the development history of the combined PCL-hBN technique for ease of understanding of the experimental technical part. The best method to fabricate a twisted homostructure is the so-called tear-and-stack technique. Unlike $CrI_3$, FGT is difficult to be exfoliated and picked up by a traditional PC transfer polymer, not to mention tearing it apart and stacking it with fine control. To address these problems, we develop a strongly adhesive dry transfer technique based on the newly invented PCL polymer[48], which can pick up many vdW magnets effectively with nearly 100 % probability under a very low operation temperature of around 60 °C.

Adopting this PCL technique to FGT led to our first attempt to access twisted FGT homostructure in 2020[49], where a plateau-like magnetoresistance was observed in this twisted FGT homostructure, even with no spacer in between. In this paper, PCL polymer gradually attaches to an FGT nanoflake, with the polymer fringe stopping in the middle of the nanoflake. Then, it picks up half of the nanoflake and drops it down on the remaining half after rotation. Although doable, there is still a low possibility of success, and it is difficult to control the whole process finely regarding the stopping fringe in the middle of FGT, preventing a further efficient and systematic investigation. More seriously, this rough tear-and-stack method can only be applied to a large-area exfoliated nanoflake, which generally hosts a large thickness, e.g., 40 nm. This, thus, would smear out the intriguing quantum effects of a twisting interface effect, if any.

The PCL method is further developed into a PCL and hBN combined technique, enabling the reliable fabrication of different twisted homostructures of diverse materials with fine control[50]. Such a new twisted fabrication method realizes the tear-and-stack process by taking advantage of the uniqueness of PCL, which is extremely sticky and can pick up the target material like FGT, for example. At the same time, it is more challenging for hBN. Most importantly, one can readily achieve fine angle control with a resolution of 0.015° and twisted fabrication on a small nanoflake of several micrometres wide and several nanometers thin. These long-lasting technique developments[48-50] eventually allow us to confidently investigate the emergent quantum phenomena in twisted metallic FGT homostructures for the first time. This new universal twisting fabrication method developed for diverse vdW magnets is extremely suitable for making twisted metallic vdW magnets such as twisted FGT homostructures.

**Transport measurement**

The substrate with twisted FGT is coated with poly (methyl methacrylate) (PMMA) through a spin coating process. A pattern of electrodes is drawn on the PMMA using electron beam lithography. For the electrodes, an electron beam evaporator is used to deposit titanium and gold, with thicknesses of 10 and 85 nm, respectively. Magnetoresistance and Hall resistance of twisted FGT are measured using the standard lock-in technique with SR830 (Stanford Research System). The device is placed inside a cryostat to maintain cryogenic temperatures. The magnetic field is applied perpendicular to the surface of the twisted FGT sample.

**Magnetic simulations**

To perform the micromagnetic simulations, we discretize the free energy in Eq. 1 on a triangular lattice with the lattice constant $A_0 = 2$nm, and the system size is taken as $L_x = 200A_0$, $L_y = 100\sqrt{3}A_0$. The intralayer DMI at different twisting angles is calculated by using the Lévy and Fert's model as detailed in the Supplementary Note 1. The other parameters are taken as $J = 2.0 \times 10^{-21}$ Joule, $J_t = 3.2 \times 10^{-23}$ Joule, $K = 1 \times 10^{-24}$ Joule, which are fitted from experimental data.

Subsequently, the magnetic ground state can be calculated by numerically minimizing the total energy, which can be implemented by different methods, such as the steepest descent. In this work, the magnetic ground state at different twist angles is obtained using VAMPIRE under the field-cool program ($B = 0.1$Tesla) and Monte Carlo integrator. During the simulation, the cooling time is set to 100 picoseconds with a time-step of 1 femtosecond, and the temperature is fixed at 20 K.

**Data availability**

Relevant data supporting the key findings of this study are available within the Article and the Supplementary Information. The raw data generated during the current study are available from the corresponding authors upon request.

**Code Availability**

The computer code used in this study is avaiailable in Code Ocean with the identifier 10.24433/CO.9659402.v1.

(2022).
62. Tai, L. X., Dai, B. Q., Li, J., Huang, H. S., Chong, S. K., Wong, K. L., Zhang, H. R., Zhang, P., Deng, P., Eckberg, C., Qiu, G., He, H. R., Wu, D., Xu, S. J., Davydov, A., Wu, R. Q. & Wang, K. L. Distinguishing the Two-Component Anomalous Hall Effect from the Topological Hall Effect. *Acs Nano* **16**, 17336–17346 (2022).
63. Hwang, I., Coak, M. J., Lee, N., Ko, D.-S., Oh, Y., Jeon, I., Son, S., Zhang, K., Kim, J. & Park, J.-G. Hard ferromagnetic van-der-Waals metal (Fe,Co)$_3$GeTe$_2$: a new platform for the study of low-dimensional magnetic quantum criticality. *Journal of Physics: Condensed Matter* **31**, 50LT01 (2019).

**Acknowledgements**

We acknowledge Suhan Son and Jingyuan Cui for helpful discussions. The work at CQM and SNU was supported by the Samsung Science & Technology Foundation (Grant No. SSTF-BA2101-05), and the Leading Researcher Program of the National Research Foundation of Korea (Grant Nos. 2020R1A3B2079375 and RS-2020-NR049405). Y.-H.L. is supported by the Fundamental Research Funds for the Central Universities, the National Natural Science Foundation of China (Grant No. 12404056) and the Natural Science Foundation of Tianjin, China (Grant No. 24JCQNJC01910). R.C. is supported by the National Science Foundation (USA) under Award No. DMR-2339315. One of the authors (J.-G.P.) acknowledges the hospitality of the Indian Institute of Science, where part of the manuscript was written through the generous support of the Infosys Foundation.

**Author contributions**

K.-X.Z. and J.-G.P. initiated and supervised the project. H.K. did all the experiments, including synthesizing the single-crystal samples, fabricating pristine and twisted devices, and conducting transport measurements with help from K.-X.Z. and G.P. K.-X.Z. and H.K. analyzed the experimental data. Y.-H.L. and R.C. performed theoretical calculations. All the authors participate in discussions. K.-X.Z., H.K., Y.-H.L. and J.-G.P. wrote the manuscript with contributions from all authors.

**Competing interests**

The authors declare no competing interests.

**Additional information**

Supplementary information: The online version contains supplementary materials available at xxx.

# Figures

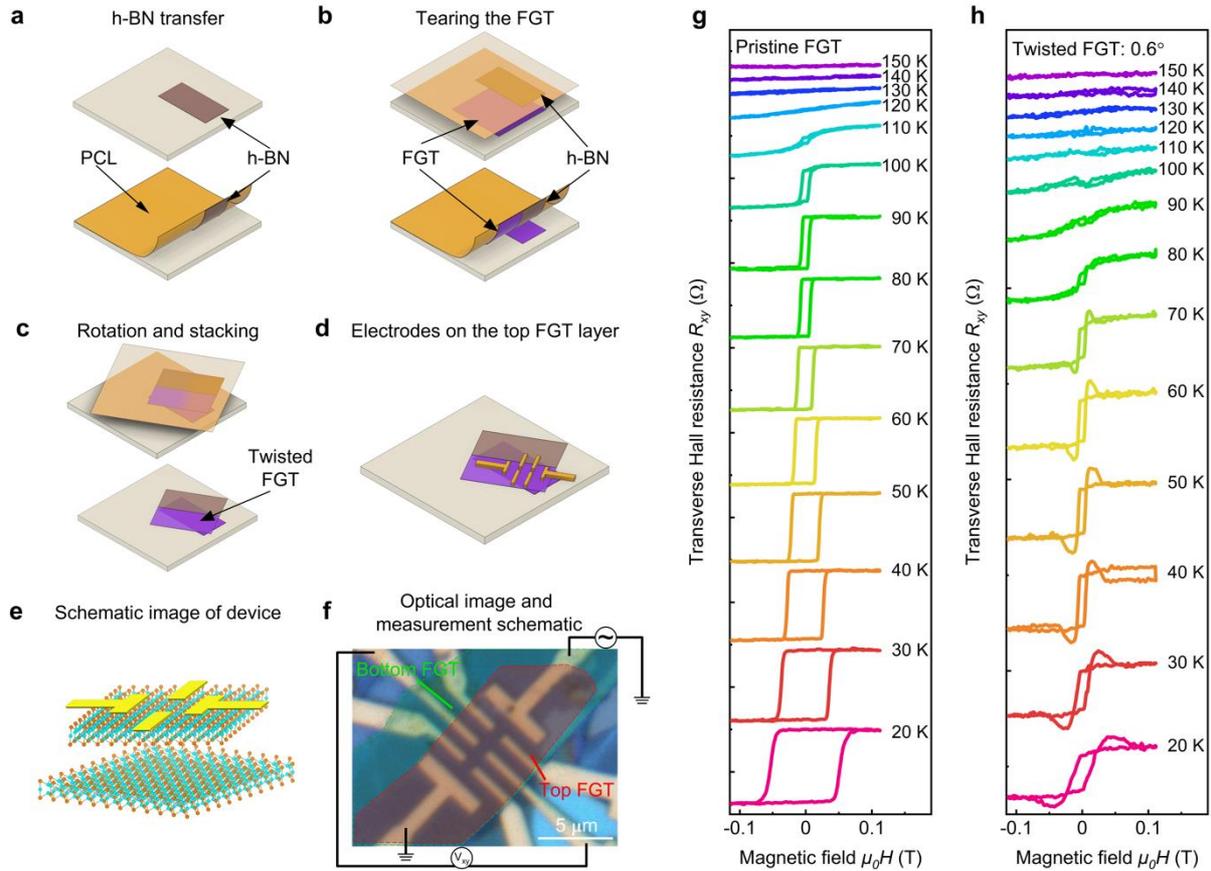

**Fig. 1 | Twisted FGT device fabrication and emergent topological Hall effect. a-d**, Illustration of twisted FGT device fabrication. There are four major steps: Pick up h-BN nanoflake by PCL method (**a**), PCL/hBN covers half of FGT nanoflake and lift to tear the FGT nanoflake (**b**), Rotate to fine control the twist angle and drop down to stack the FGT/FGT homostructure (**c**), Hall-bar geometry electrodes are patterned onto the twisted region (**d**). **e**, Schematic of a twisted FGT homostructure. **f**, Typical optical image of a real twisted FGT device. The white scale bar is 5 μm. The Hall electrodes are patterned onto the twisted overlapping region. **g**, Transverse Hall resistance $R_{xy}$ as a function of an out-of-plane magnetic field $H$. It shows the typical ferromagnetic hysteresis loops at low temperatures. **h**, $R_{xy}$ versus magnetic field $H$ for a twisted FGT device at a twist angle of 0.6°. At low temperatures, two asymmetric $R_{xy}$ hump unexpectedly appear at both positive and negative fields in addition to the anomalous Hall loop, manifesting an emergent topological Hall effect by twisting.

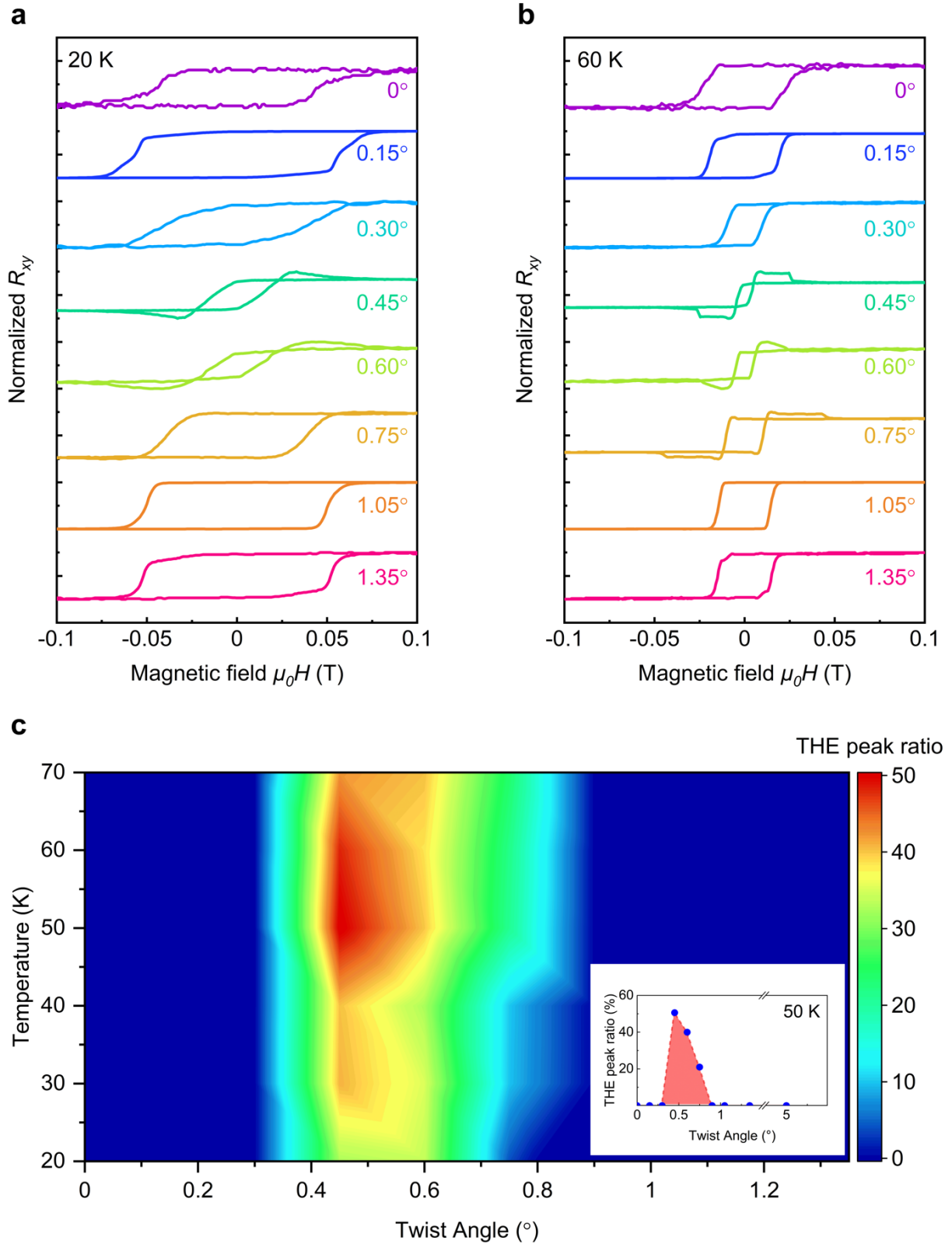

**Fig. 2 | Twist angle dependence of the new topological Hall effect. a**, Normalized $R_{xy}$ as a function of magnetic field $H$ under 20 K for different twisted FGT samples with individual twist angles. Topological Hall humps exist for twist angles between 0.45° and 0.75° in addition to the anomalous Hall loops. The other twisted FGT samples only have the typical anomalous Hall loops as that of single pristine FGT nanoflakes. **b**, Similar $R_{xy}$ versus $H$ curves at different

twist angles, but under the temperature of 60 K. The twist angle dependence of the new topological Hall effect exhibits universality at all low temperatures listed above. **c**, The colour map of the THE Peak ratio [defined as (topological Hall hump) / (topological Hall hump + anomalous Hall magnitude) × 100%] in the temperature-twist angle phase space. The inset indicates that only twist angles of 0.45°, 0.6°, and 0.75° can produce the topological Hall effect among the whole twist angle range from 0° to 5°, large enough compared to the previous twisted case for vdW magnetic insulators[42-44]. The colour map shows the magic angles regime between 0.45° and 0.75°, where the topological Hall effect emerges. Meanwhile, these results also demonstrate reliable reproducibility for different temperatures and individual twisted devices. Note that the THE ratio can reach about 50 %, showing a colossal topological Hall effect comparable to its intrinsic gigantic AHE governed by large Berry curvatures of band topology.

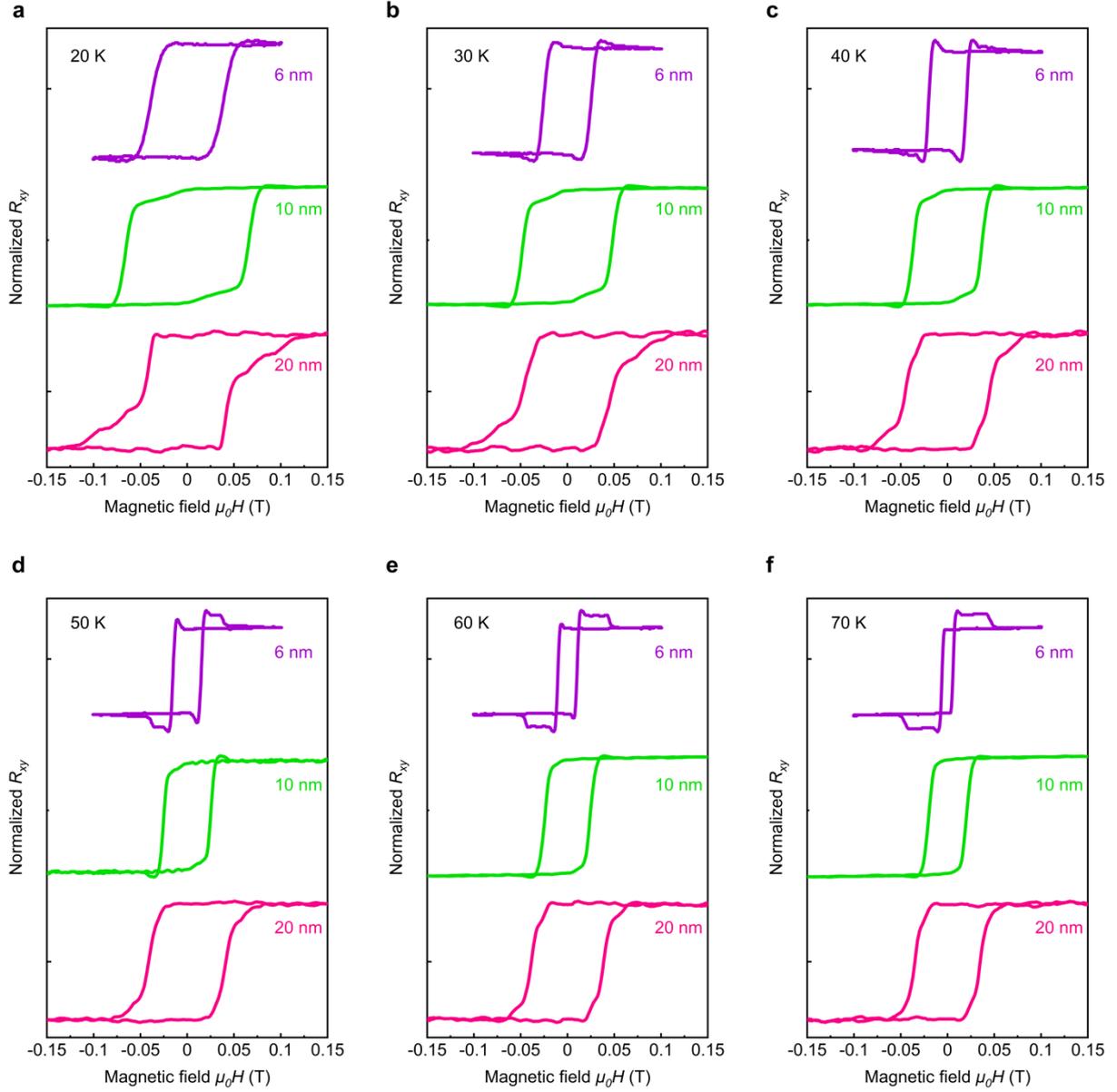

**Fig. 3 | Thickness dependence of the twisting-induced topological Hall effect. d**, Normalized $R_{xy}$ versus magnetic field $H$ under 50 K for 0.75° twisted FGT devices of different thicknesses. The topological Hall loop is significant for a ~6 nm-thick twisted device, suppressed for a ~10 nm-thick twisted device and vanishing for a ~20 nm-thick twisted device. **a-c,e-f**, Similar results as in (**d**), but under the temperature of 20, 30, 40, 60, and 70 K. Considering that the out-of-plane magnetic interaction expands to ~5 nm in FGT, this suppression upon increasing thickness underlines the interfacial effect at the twisting interface, for accounting for the artificially created topological Hall effect by twisting.

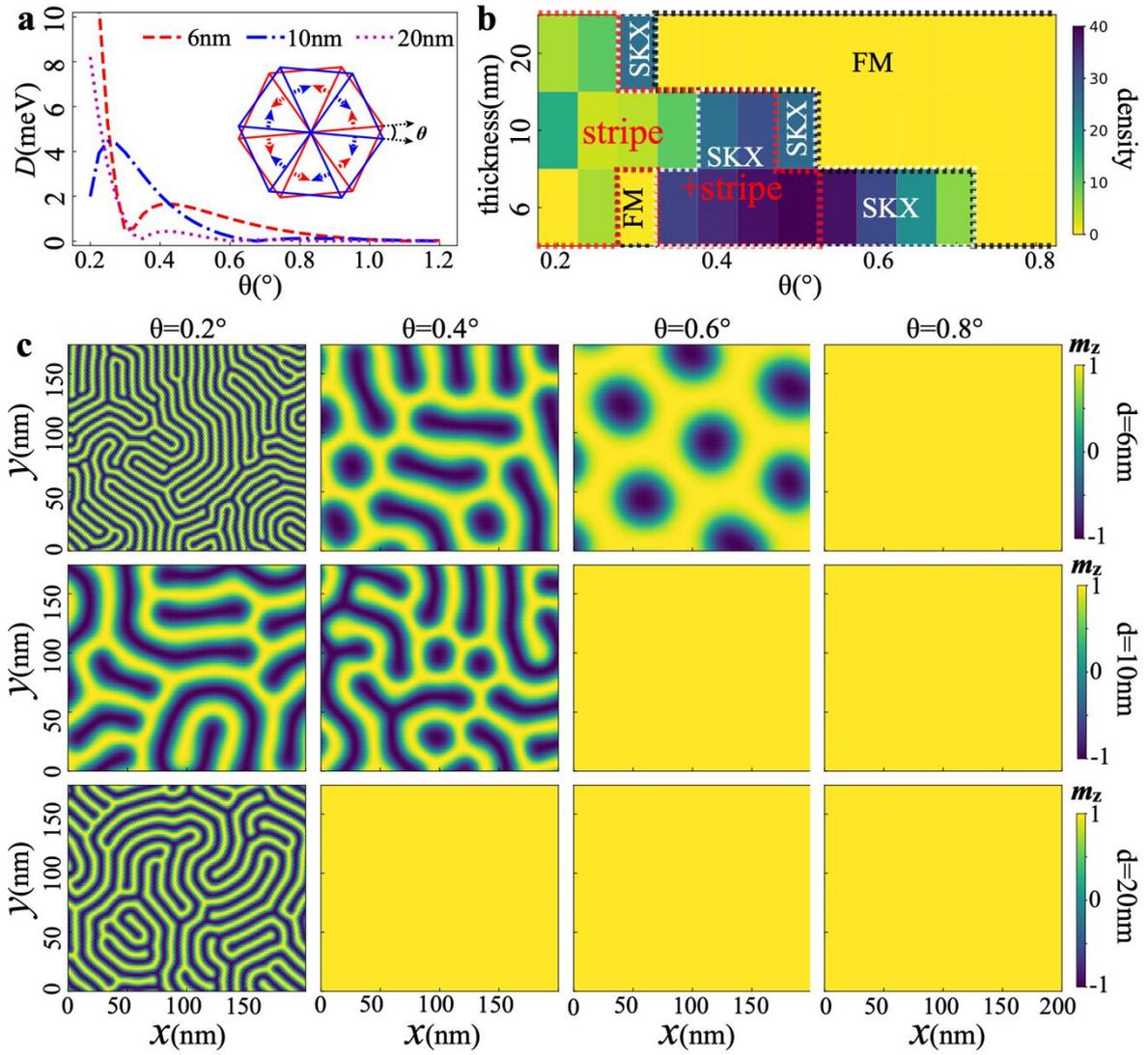

**Fig. 4 | Micromagnetic simulation for FGT twisted system. a,** Magnitude of the intralayer DMI $D$ as a function of twisting angle $\theta$ for three different thicknesses 6 nm, 10 nm, and 20 nm. The inset illustrates the direction of the intralayer DMI on top (blue) and bottom (red) layer. **b,** Phase diagram represented by the skyrmion density inside the entire simulated region versus twisting angle $\theta$ and thickness. Here, SKX, FM, and stripe refer to the skyrmion lattice, ferromagnetic state, and stripe state, respectively. **c,** Spin configurations of the bottom layer represented by the $z$-component of the magnetization ($m_z$) at each site for representative twisting angle $\theta$ (labeled on the top) and thickness (labeled on the right side).

# Supplementary Information for:

## Emergent giant topological Hall effect in twisted Fe$_3$GeTe$_2$ metallic system

Hyuncheol Kim[1,2,3†], Kai-Xuan Zhang[1,2,3†*$], Yu-Hang Li[4†], Giung Park[1,2,3], Ran Cheng[5,6], Je-Geun Park[1,2,3*]

[1]Department of Physics and Astronomy, Seoul National University, Seoul 08826, South Korea
[2]Center for Quantum Materials, Department of Physics and Astronomy, Seoul National University, Seoul 08826, South Korea
[3]Institute of Applied Physics, Seoul National University, Seoul 08826, South Korea
[4]School of Physics, Nankai University, Tianjin 300071, China
[5]Department of Electrical and Computer Engineering, University of California, Riverside, CA 92521, USA
[6]Department of Physics and Astronomy, University of California, Riverside, CA 92521, USA
[$]Present address: Department of Physics, Washington University in St. Louis, St. Louis, Missouri 63130, USA

[†] These authors contributed equally to this work
[*] Corresponding authors: Kai-Xuan Zhang (email: kxzhang.research@gmail.com), and Je-Geun Park (email: jgpark10@snu.ac.kr)

**This PDF file includes:**

   Materials and Methods

   Supplementary Notes 1 to 8

   Supplementary Figures S1 to S14

   Supplementary References

**Materials and Methods**

Preparation of polycaprolactone solution as adhesive polymer

Polycaprolactone (PCL, Sigma Aldrich, average $M_n$ = 80,000) was utilized as an adhesive polymer for a tear-and-stack method. To prepare the adhesive PCL stamp, PCL was dissolved in stabilized tetrahydrofuran (THF, Sigma-Aldrich) to a concentration of 15% by mass ratio. The process was initiated by affixing polydimethylsiloxane (PDMS) to a slide glass and securing it with transparent tape (Scotch Crystal Tape) (Fig. S1a, b). The PCL solution was then cast onto the PDMS region, and a uniform thin PCL film was formed through spin-coating at 3,000 rpm for 20 seconds (Fig. S1c, d). To optimize the film for adhesion, the slide glass with PCL film was subsequently annealed at 90℃ for 50 min, which effectively reduced its surface roughness.

Thickness of twisted FGT devices

To address the thickness dependence of the topological Hall effect (THE) in twisted FGT, we made three twisted FGT devices with the same twist-angle of 0.75°, using three individual FGT nanoflakes of different thicknesses: 6, 10, and 20 nm. Using this controlled thickness experiment, we confirmed that the magnitude of THE in twisted FGT systems is suppressed by increasing the thickness of the FGT nanoflake. For example, the THE phenomena in twisted FGT systems are observable and almost comparable for FGT nanoflakes equal to or thinner than 10 nm. To investigate the twist-angle dependence of THE, we fabricated a total of 10 twisted FGT devices using the FGT nanoflakes with thicknesses of about 6~8 nm.

Supplementary Note 1. Calculation of Dzyaloshinskii-Moriya interaction

Twisting only suppresses the band dispersion near K points, while it does not alter the band structure near the Gamma point[1]. Since FGT is a metal whose band, originating from the 3d orbitals of Fe I and Fe II, as well as p orbitals of Ge and Te atoms, spreads across the entire Brillouin zone instead of being confined near the K point, the twisted FGT system remains metallic with its Fermi wavelength strongly suppressed. Consequently, we can describe the twisted bilayer FGT system by using the Hamiltonian

$$H = H_0 + \Gamma_{ex}\vec{s}\cdot\vec{S}_A + \Gamma_{ex}\vec{s}\cdot\vec{S}_B + \lambda_d\vec{s}\cdot\vec{l}, \quad \text{S 1}$$

where the first term $H_0$ is the Hamiltonian for an itinerant electron with spin $\vec{s}$, the second and third terms are the sd exchange interaction between the itinerant electron and local spin $\vec{S}_{A/B}$ with $\Gamma_{ex}$ its strength, the third term is the spin-orbit coupling. In the framework of the Lévy-Fert model, the DMI can be derived as a second-order perturbation correction to the total energy:

$$E_2 = \sum_{k\neq k'} f(k) \frac{\langle k\sigma|\Gamma_{ex}\vec{s}\cdot\vec{S}_A|k'\sigma'\rangle_\lambda \langle k'\sigma'|\Gamma_{ex}\vec{s}\cdot\vec{S}_B|k\sigma\rangle_\lambda}{E_k - E_{k'}}, \quad \text{S 2}$$

where $|k\sigma\rangle_\lambda = |k\sigma\rangle + \sum_{k'}\langle k'\sigma'|\lambda_d\vec{s}\cdot\vec{l}|k\sigma\rangle/(E_k - E_{k'})|k'\sigma'\rangle$ is the first-order perturbative eigenstate with $|k\sigma\rangle$ ($E_k$) unperturbed eigenstate (eigenenergy). Substituting the first-order perturbative eigenstate into Eq. (S2) and using the relation $\sum_{\sigma_1\sigma_2\sigma_3}\langle\sigma_1|\vec{s}\cdot\vec{S}_A|\sigma_2\rangle\langle\sigma_2|\vec{s}|\sigma_3\rangle\langle\sigma_3|\vec{s}\cdot\vec{S}_B|\sigma_1\rangle = -i/4(\vec{S}_A \times \vec{S}_B)$, we finally obtain the expression of DMI

$$\vec{D}_{ijk}(\vec{R}_{ki}, \vec{R}_{kj}, \vec{R}_{ij}) = -V_1 \frac{\sin[k_F(|\vec{R}_{ki}|+|\vec{R}_{kj}|+|\vec{R}_{ij}|)+\left(\frac{\pi}{10}\right)Z_d](\vec{R}_{ki}\cdot\vec{R}_{kj})(\vec{R}_{ki}\times\vec{R}_{kj})}{|\vec{R}_{ki}|^3|\vec{R}_{kj}|^3|\vec{R}_{ij}|}, \quad \text{S 3}$$

where $\vec{R}_{ki}$, $\vec{R}_{kj}$ and $\vec{R}_{ij}$ are vectors connecting corresponding sites, $V_1 = [135\pi/32][(\lambda_d\Gamma_{ex}^2)/(E_F^2 k_F^3)]\sin(Z_d\pi/10)$ is the material parameter defining the DMI strength.

Here, $E_F$ and $k_F$ are the Fermi energy and Fermi wavelength, $\lambda_d$ is the spin-orbital coupling parameter, $\Gamma_{ex}$ is the parameter characterizing the exchange interaction between localized spins and the spins of conduction electrons, and $Z_d$ denotes the number of d-orbital electrons in Fe atoms. Summing over all sites $k$ gives the DMI $\vec{D}_{ij}$ between two sites $i$ and $j$.

In twisted bilayer FGT, $\vec{D}_{ij}$ can be calculated involving the twisting lattice between neighboring layers. Besides, as the contribution to $\vec{D}_{ij}$ from site $k$ that is far from sites $i$ and $j$ is negligible, a proper cut is necessary when summing over $k$. Note that both intralayer and interlayer DMIs are allowed, although the interlayer DMI vanishes at the AA point in the moiré lattice. For simplicity, we neglect the difference between Fe I and Fe II inside FGT, and assume a triangular lattice with a lattice constant $a_0 = 4$ Å as shown in Fig.S11(a). Remarkably, we find that, despite the spatial variation, the interlayer DMI is about four orders of magnitude smaller than the intralayer DMI. In addition, as schematically shown in Fig.S12(b), the intralyer DMI at different layers exhibits a chirality obeying the relation $\vec{D}_{ij} = D_0(-1)^l \hat{z} \times \hat{e}_{ij}$ where $l = 1$ or $2$ refers to the top or bottom layer. On the other hand, its magnitude is universal across the entire moiré lattice, which decreases sharply as the twisting angle $\theta$ [Fig.S11(b)].

## Supplementary Note 2. Continuum model of twisted bilayer FGT

In the continuum limit, twisted bilayer FGT can generally be described by the free energy

$$H = \int d\vec{r}^2 \{\sum_l [Ja_0^2(\nabla \vec{m}_l)^2 - K(m_l^z)^2] + J_t(r)\vec{m}_1 \cdot \vec{m}_2 + \vec{D}_t(r) \cdot \vec{m}_1 \times \vec{m}_2$$
$$+ D_0(\vec{r}) a_0 \sum_l (-1)^l [\hat{y} \cdot (\vec{m}_l \times \partial_x \vec{m}_l) - \hat{x} \cdot (\vec{m}_l \times \partial_y \vec{m}_l)]\}, \quad \text{S 4}$$

where $Ja_0^2$ is the exchange stiffness, $K$ is the uniaxial anisotropy, $J_t(\vec{r}) =$

$J_t \sum_{\alpha=1}^{3} \cos(\vec{q}_\alpha \cdot \vec{r})$ denotes the spatial variable interlayer exchange interaction, $\vec{D}_t(\vec{r})$ is the interlayer DMI, and the last term refers to the intralayer DMI. Our calculation of the DMI shows that the interlayer DMI is about four orders of magnitude smaller than the intralayer DMI, which is also smaller than the exchange stiffness $Ja_0^2$ as well as the anisotropy $K$. Consequently, this term can be readily neglected. On the other hand, FGT is a van der Walls material whose interlayer exchange interaction is much weaker than others, especially in twisted bilayer FGT, the layer distance of which can even be nm. Furthermore, as shown in Fig. S12, different $J_t$ does not result in any quantitative difference in our simulations. We can thus further neglect the spatial variation of $J_t(\vec{r})$ and replace it with a uniform value. Therefore, the continuum energy of twisted bilayer FGT finally takes the form

$$H = \int d\vec{r}^2 \{\sum_l [Ja_0^2(\nabla \vec{m}_l)^2 - K(m_l^z)^2] + J_t \vec{m}_1 \cdot \vec{m}_2 \\ +D(\theta)a_0 \sum_l (-1)^l [\hat{y} \cdot (\vec{m}_l \times \partial_x \vec{m}_l) - \hat{x} \cdot (\vec{m}_l \times \partial_y \vec{m}_l)]\}. \quad \text{S 5}$$

The parameters in Eq.S3 can be obtained from the experimental data. In the Stoner-Wolfarth model, the coercive field $\mu_0 H_c$ is determined by the uniaxial magnetic anisotropy through the relation $\mu_0 H_c M = K$, where $M$ is the saturation magnetization. The exchange stiffness can then be fit from the Curie temperature in the framework of the Curie-Weiss theory through $3k_B T_c = 2S(S+1)(\sum_{\langle ij \rangle} J_{ij} + K)/N$, where $S = 5/2$ is the spin of the magnetic Fe atom, $N$ is the number of nearest neighbor sites, and the summation is performed over the exchange interaction between all nearest neighbors. Note that we have set that the spins of both Fe I and Fe II are $S = 5/2$. Consequently, the anisotropy $K$ and the exchange stiffness fitted from the experimental data are $K = 1.0 \times 10^{-24}$ Joule, and $J = 2.0 \times 10^{-21}$ Joule. The interlayer exchange interaction is taken as $J_t = 3.2 \times 10^{-23}$ Joule.

Supplementary Note 3. Characterization of material properties

The fundamental physical properties of the FGT were characterized (Fig. S13). An optical image of the FGT single crystal, along with its XRD and EDS data, is presented in Fig. S13 a. The temperature-dependent resistance curves for both an isolated FGT nanoflake and the FGT homostructure device are shown in Fig. S13 b and c, respectively. These results ensured the high crystalline quality and the typical behavior of FGT consistent with previous reports.

Supplementary Note 4. Hysteresis loop along out-of-plane and in-plane magnetic field direction

As shown in Fig. S14, the FGT sample used in this work still exhibits a clear hysteresis loop when the magnetic field is directed along the in-plane direction, demonstrating a normal perpendicular ferromagnetism as reported in previous literature[3-5]. It ought to be noted that the topological Hall effect only emerges in the twisted FGT/FGT devices within the magic angle region, and has never been observed in isolated FGT devices. This strongly supports our conclusion that our central finding of the topological Hall effect is indeed due to the twisting effect rather than trivial reasons.

Supplementary Note 5. Temperature-dependent topological Hall effect in twisted FGT

The topological Hall effect (THE) is often more pronounced at lower temperatures. However, there are notable exceptions in which THE becomes stronger at elevated temperatures, depending on the material system. One such example is the α-phase of MnSi[6]. The magnitude of THE in MnSi increases slightly with temperature within a certain range. Furthermore, the temperature window over which the α-phase exhibits a topological Hall signal becomes broader at higher temperatures.

These observations suggest that the temperature dependence of THE is highly material-specific and can be influenced by several factors such as magnetic anisotropy, exchange interactions, and thermal stability of the topological spin textures. In our study, the twisted FGT system exhibits a similar trend to MnSi, where the THE becomes more prominent at slightly higher temperatures, reinforcing the idea that this behavior is not universal but rather material-dependent.

In particular for twisted FGT, the formation of topological spin textures is governed by the interplay between magnetic anisotropy, exchange interaction, and the Dzyaloshinskii–Moriya interaction (DMI). When electrons move around this topological spin texture, they acquire an additional phase, leading to the topological Hall effect.

It is well known that the magnetic anisotropy of ferromagnet FGT is large at low temperatures but decreases with rising temperature. At low temperatures, the ferromagnetic ground state of FGT is more stable due to the dominance of strong magnetic anisotropy, which suppresses the emergence of topological spin textures. In such a regime, the DMI is typically insufficient to overcome the energy barrier required to stabilize topological spin textures, thereby reducing the likelihood of observing a topological Hall effect. At relatively higher temperatures, magnetic anisotropy has been reduced, and thus the topological Hall effect is present due to the comparatively stronger DMI. However, at extremely high temperatures close to the Curie temperature, all magnetic ground states and interactions are greatly destroyed. Thus, the corresponding topological spin texture and topological Hall effect vanish. Our magnetic simulation results further support this picture, indicating that skyrmion formation becomes energetically favorable at specific twist angles, which highlights the importance of competitive magnetic interactions at play. These consistent results between theory and experiments highlight the importance of both geometrical and interaction-based tuning parameters in stabilizing topological spin textures in twisted FGT systems.

Supplementary Note 6. Skyrmion size in twisted FGT

The skyrmion size can be estimated from the magnitude of the topological Hall resistivity[7-9]. This estimation is based on the following relation:

$$\boldsymbol{\rho_{THE} = R_0 P \Phi_0 N_{sk}},  \quad \text{(Eq. S6)}$$

Where the $\rho_{THE}$ is the topological Hall resistivity, $R_0$ is the ordinary Hall coefficient, $P$ is the spin polarization, $\Phi_0$ is the magnetic flux quantum, and $N_{sk}$ is the skyrmion density.

In our study, the magnitude of $\rho_{THE}$ in twisted FGT with twist angle 0.45° at 60 K is as large as 0.71 µΩ·cm. The ordinary Hall coefficient, $R_0$, was determined from the linear slope of the Hall resistivity data in the high-field regime where the spins are fully saturated. The value is 1.51×10$^{-9}$ m³/C at 60 K. The spin polarization, $P$, was inferred as 0.53 at 60 K from the previous report[10]. Based on these physical parameters and Eq. S6, the estimated skyrmion density is approximately 2.15×10$^{15}$ m$^{-2}$, corresponding to a skyrmion diameter of ~23 nm. In addition, we compared the experimentally estimated skyrmion size with the results obtained from our simulation. For a twist angle of 0.45°, the skyrmion size from the simulation is about 35 nm. This value agrees quantitatively with the previously estimated skyrmion diameter, which is approximately 23 nm.

Supplementary Note 7. Mean free path and skyrmion size

To validate the interpretation of the THE as originating from the adiabatic evolution associated with scalar spin chirality, such as that arising from topological spin textures, it requires a skyrmion diameter to be much larger than the electronic mean free path. The electron mean free path was estimated using the Drude model, in which the mean free path $l$ is given by the product of the Fermi velocity $v_F$ and the scattering time $\tau$. While the exact value of $v_F$ for FGT is not well established in the literature, we estimated it based on the ARPES data[11]. According to Fig. 2 of that work, three energy bands—labelled α, γ, and δ—are observed near the Fermi surface. By extracting the approximate slopes of these bands from Fig. 2(c), we estimate the Fermi velocity to be approximately $3.04×10^5$, $7.60×10^5$, and $6.08×10^5$ m/s for α, γ, and δ bands, respectively. The final scattering time $\tau$ was calculated using the relation:

$$\tau = \frac{m^*}{ne^2\rho}, \quad \text{(Eq. S7)}$$

Where $m^*$ is the effective mass (taken as the rest mass of the electron $m_e$), $n$ is the carrier density, $e$ is the elementary charge, $\rho$ is the resistivity. The Hall coefficient measured for the twisted FGT device with 0.45° twisted angle is $1.51×10^{-9}$ m³/C, yielding a carrier density n ~ $4.14×10^{27}$ m$^{-3}$. With a measured resistivity of 434 μΩ·cm, the resulting scattering time is $1.98×10^{-15}$ s.

Consequently, the mean free paths for electrons in the α, γ, and δ bands are estimated to be 0.60, 1.50, and 1.20 nm, respectively. These values indicate that the electron mean free path in FGT is ~ 1 nm, which is smaller than the previously estimated skyrmion diameter. This calculation validates our interpretation of THE in terms of an effective magnetic field from real-space chiral spin textures.

Supplementary Note 8. Discussions on the cases where the "two-channel AHE" scenario is difficult to be straightforwardly applicable in our twisted FGT devices.

We acknowledge that the recently developed "two-channel Hall effect" is an important addition to the topological Hall effect, accounting for the extra hump found in certain materials. We carefully reviewed the references[12-14], and appreciate their scientific discoveries and discussions on the two-channel anomalous Hall components with opposite signs. However, this "two-channel" scenario significantly differs from the conditions found in our present study based on several important experimental facts as follows:

(1) Device fabrication perspective: First, our work focuses on FGT homostructure, not heterostructures. Specifically, we employed a tear-and-stack method, in which a single FGT nanoflake is split and stacked to form the twisted double-layer device. This ensures that only one intrinsic type of anomalous Hall response is present in the system. Moreover, as shown in the Supplementary Information, the top and bottom FGT layers in our device have identical thicknesses. Consequently, the coercive fields of both layers are effectively the same. In such a case, even if a two-component AHE mechanism were hypothetically present, the resulting hysteresis would not produce pronounced hump or dip features, as the switching fields would overlap. Therefore, the two-component AHE scenario is unlikely to apply to our device configuration.

(2) THE angle dependence perspective: Second, THE only emerges within the small magic angle region with twisting angle from 0.45° to 0.75°. Outside this small twist angle region, the $R_{xy}$-H curves of the twisting device only show normal AHE loops with relatively uniform switching fields. A two-component AHE model cannot account for such a striking angle dependence of THE observed in our twisted FGT samples. If the hump and dip structures originated solely from competing AHE contributions, similar features should appear at various twist angles regardless of the moiré geometry. In contrast, our measurements show that THE features are strongly dependent on the twist angle, consistent with a twist-induced magnetic texture rather than overlapping AHE signals.

(3) Thickness and temperature dependence: Our experimental data show that THE exhibits a clear thickness and temperature dependence, reflecting the competition between interfacial magnetism and magnetic interactions. These features are consistent with theoretical analyses and cannot be explained by an accidental "two-channel" scenario.

(4) Previous FGT investigations: The two-channel scenario requires two hysteresis loops of "opposite sign" for the material system. However, according to numerous FGT literatures and over 100 FGT devices we have so far measured in our lab (FGT nanotransport is a mandatory basic training for any new student entering our lab), "opposite-signed" hysteresis has never been observed for numerous diverse FGT nanoflake devices. Moreover, our twisted FGT devices, where the two FGT layers originate from the same single nanoflake, further exclude the "two-channel" scenario.

In summary, although we respectfully agree that the two-channel scenario is an important correction, it depends on the material or device system. Several experimental facts mentioned above indicate that the "two-channel" scenario is unlikely to exist dominantly in twisted FGT devices and cannot explain the THE in our system.

**Supplementary Figures**

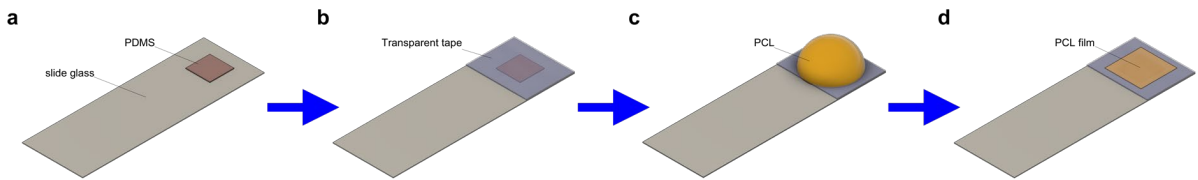

**Fig. S1 | Fabrication of PCL stamp. a** A PDMS block is attached to a slide glass. **b** The PDMS block is secured to the slide with transparent tape. **c** A PCL solution, which is the mixture of THF and PCL with 15% mass ratio, is dispensed onto the transparent tape covering the PDMS block until the PDMS surface is fully covered. **d** A uniform PCL film is formed by spin-coating at 3,000 rpm for 20 s, followed by thermal annealing at 90℃ for 50 min.

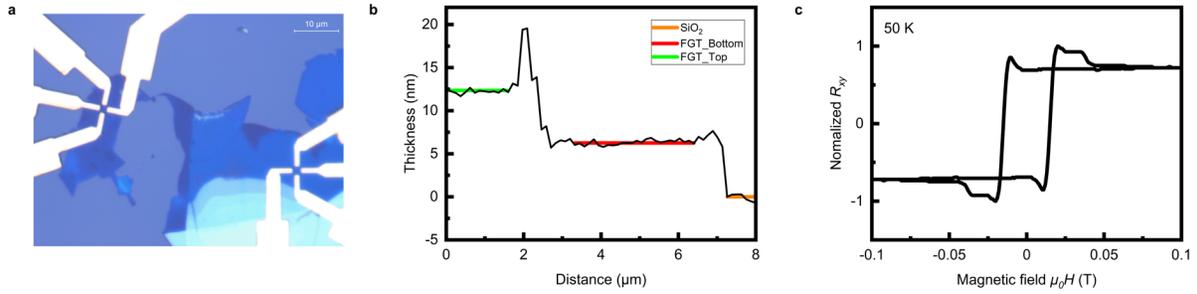

**Fig. S2 | Basic characterization of the twisted FGT device. a** An optical image of the twisted FGT device. The twist angle is 0.75°. **b** The device's thickness is about 6 nm, measured by AFM. **c** Transverse Hall resistance of the device at 50 K, as a function of magnetic field.

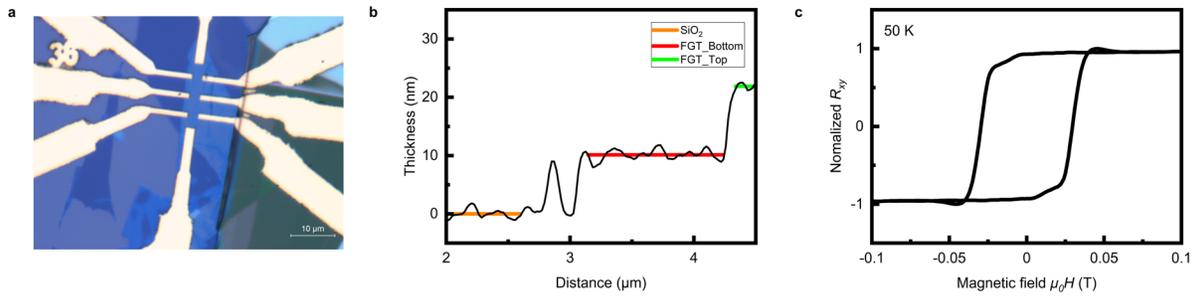

**Fig. S3 | Basic characterization of the twisted FGT device. a** An optical image of the twisted FGT device. The twist angle is 0.75°. **b** The device's thickness is about 10 nm, measured by AFM. **c** Transverse Hall resistance of the device at 50 K, as a function of magnetic field.

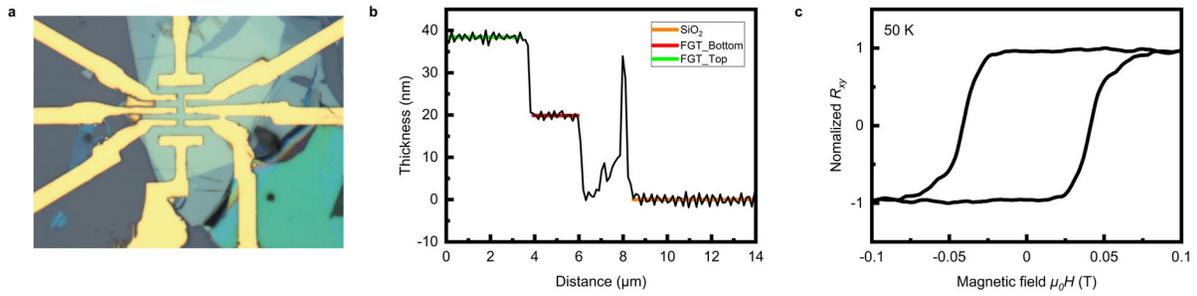

**Fig. S4 | Basic characterization of the twisted FGT device. a** An optical image of the twisted FGT device. The twist angle is 0.75°. **b** The device's thickness is about 20 nm, measured by AFM. **c** Transverse Hall resistance of the device at 50 K, as a function of magnetic field.

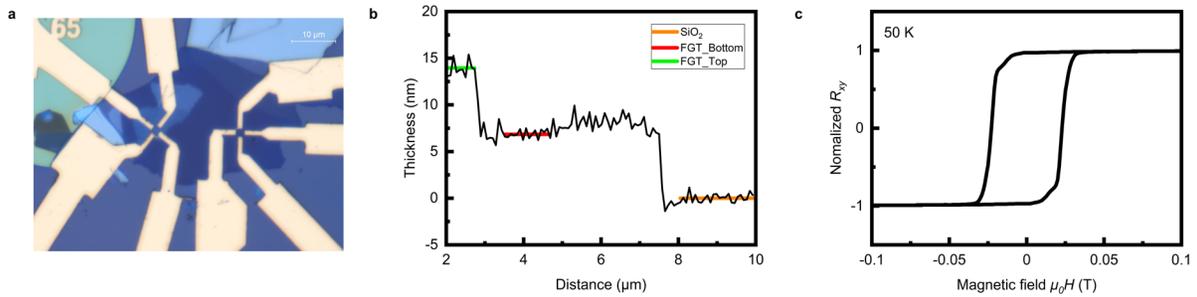

**Fig. S5 | Basic characterization of the twisted FGT device. a** An optical image of the twisted FGT device. The twist angle is 0.15°. **b** The device's thickness is about 7 nm, measured by AFM. **c** Transverse Hall resistance of the device at 50 K, as a function of magnetic field.

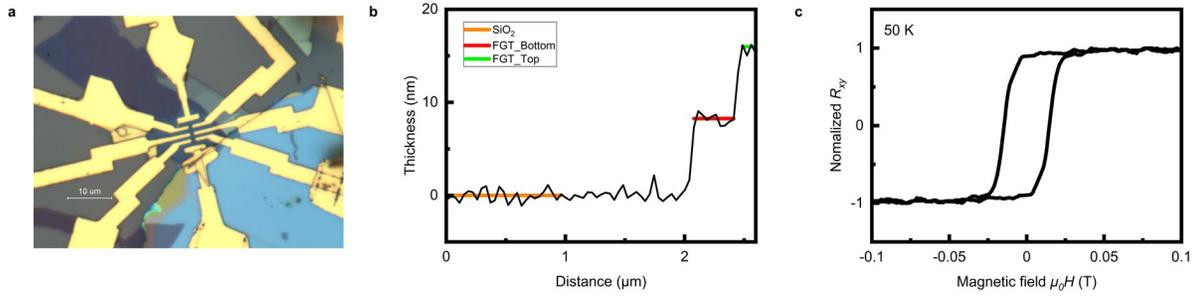

**Fig. S6 | Basic characterization of the twisted FGT device. a** An optical image of the twisted FGT device. The twist angle is 0.3°. **b** The device's thickness is about 8 nm, measured by AFM. **c** Transverse Hall resistance of the device at 50 K, as a function of magnetic field.

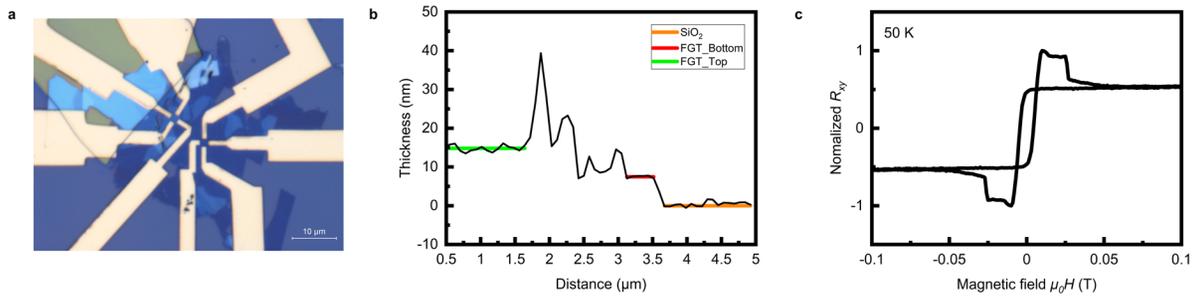

**Fig. S7 | Basic characterization of the twisted FGT device. a** An optical image of the twisted FGT device. The twist angle is 0.45°. **b** The device's thickness is about 7.5 nm, measured by AFM. **c** Transverse Hall resistance of the device at 50 K, as a function of magnetic field.

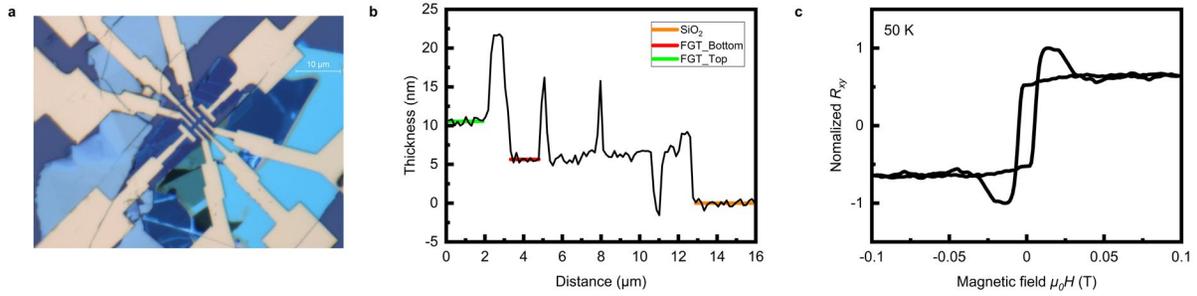

**Fig. S8 | Basic characterization of the twisted FGT device. a** An optical image of the twisted FGT device. The twist angle is 0.6°. **b** The device's thickness is about 5.5 nm, measured by AFM. **c** Transverse Hall resistance of the device at 50 K, as a function of magnetic field.

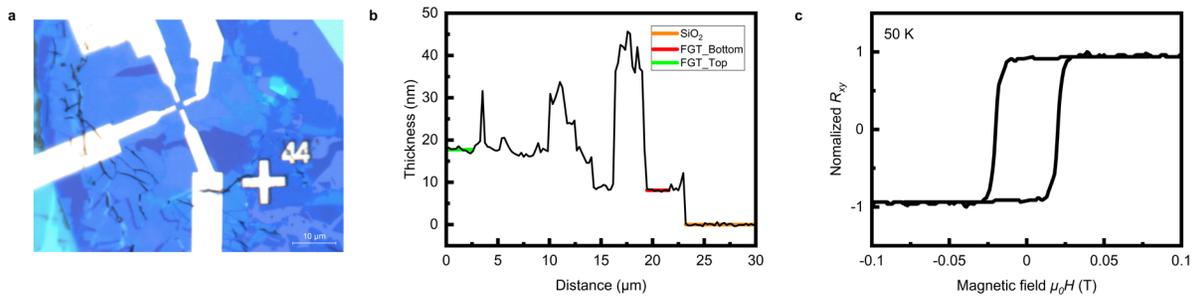

**Fig. S9 | Basic characterization of the twisted FGT device. a** An optical image of the twisted FGT device. The twist angle is 1.35°. **b** The device's thickness is about 8 nm, measured by AFM. **c** Transverse Hall resistance of the device at 50 K, as a function of magnetic field.

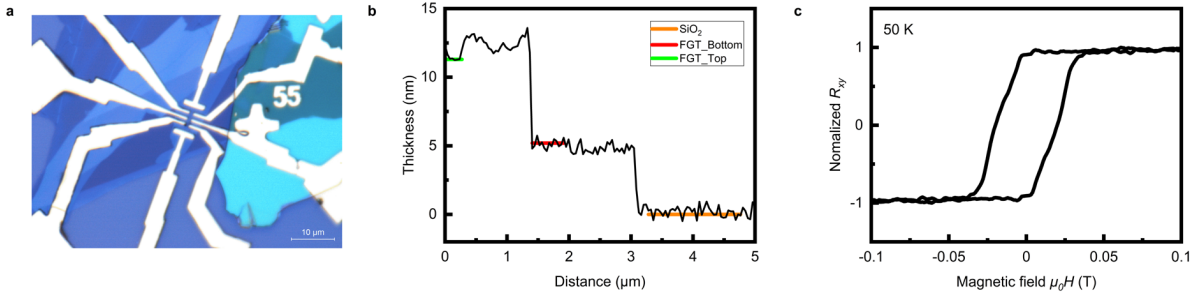

**Fig. S10 | Basic characterization of the twisted FGT device. a** An optical image of the twisted FGT device. The twist angle is 5°. **b** The device's thickness is about 5.5 nm, measured by AFM. **c** Transverse Hall resistance of the device at 50 K, as a function of magnetic field.

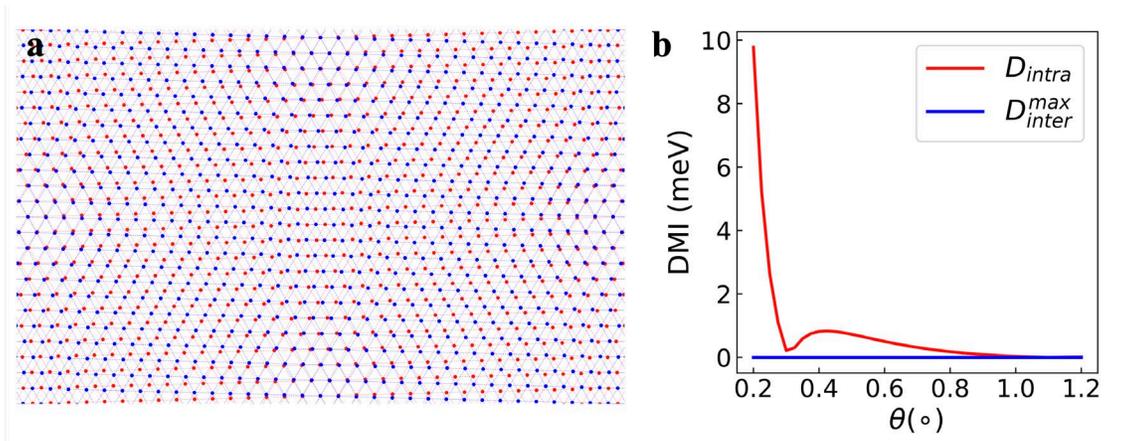

**Fig. S11 | Calculation of DMI in Twisted FGT. a** Twisted bilayer FGT moiré lattice at twisting angle $\theta = 5°$. The red lattice refers to the bottom layer, while the blue lattice refers to the top layer. **b** Magnitudes of intralayer DMI and the maximum interlayer DMI versus $\theta$. Here, $E_F = 5$ eV, $\lambda_d = 10$ meV, $\Gamma_{ex} = 20$ meV, $Z_d = 5$, and the interlayer distance is taken as $d = 6$ nm.

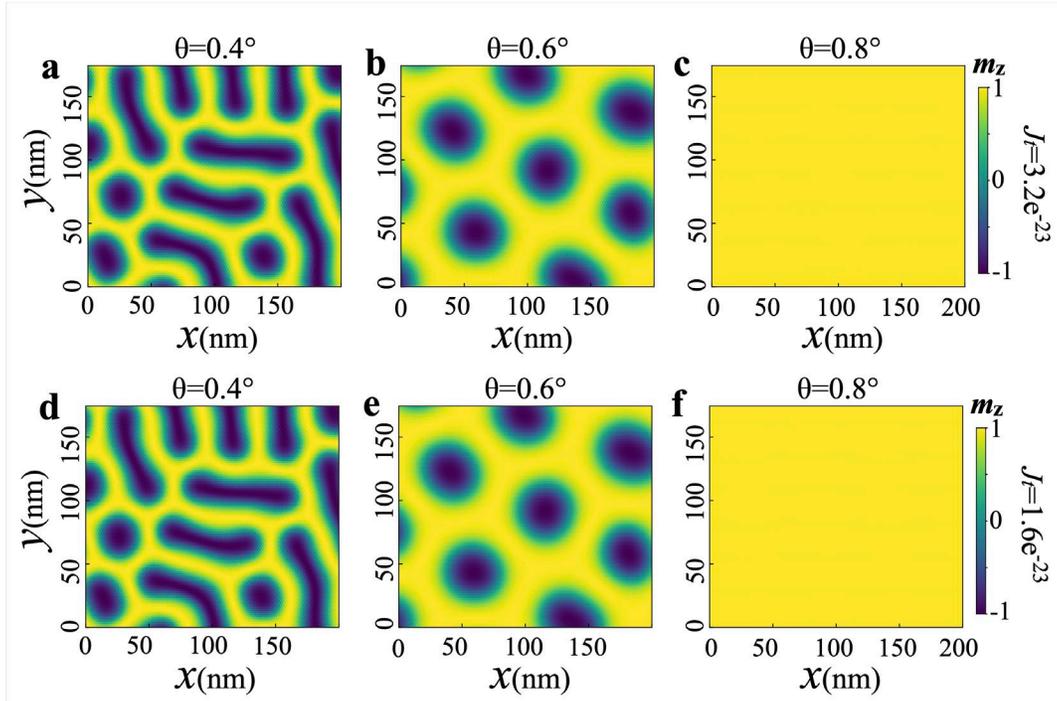

**Fig. S12 | Spin configurations at different twisting angles with different interlayer exchange interactions.** $J_t = 0.2$ meV in the top panels (**a-c**) while $J_t = 0.1$ meV in the bottom panels (**d-f**). Other parameters are the same as those in the main text.

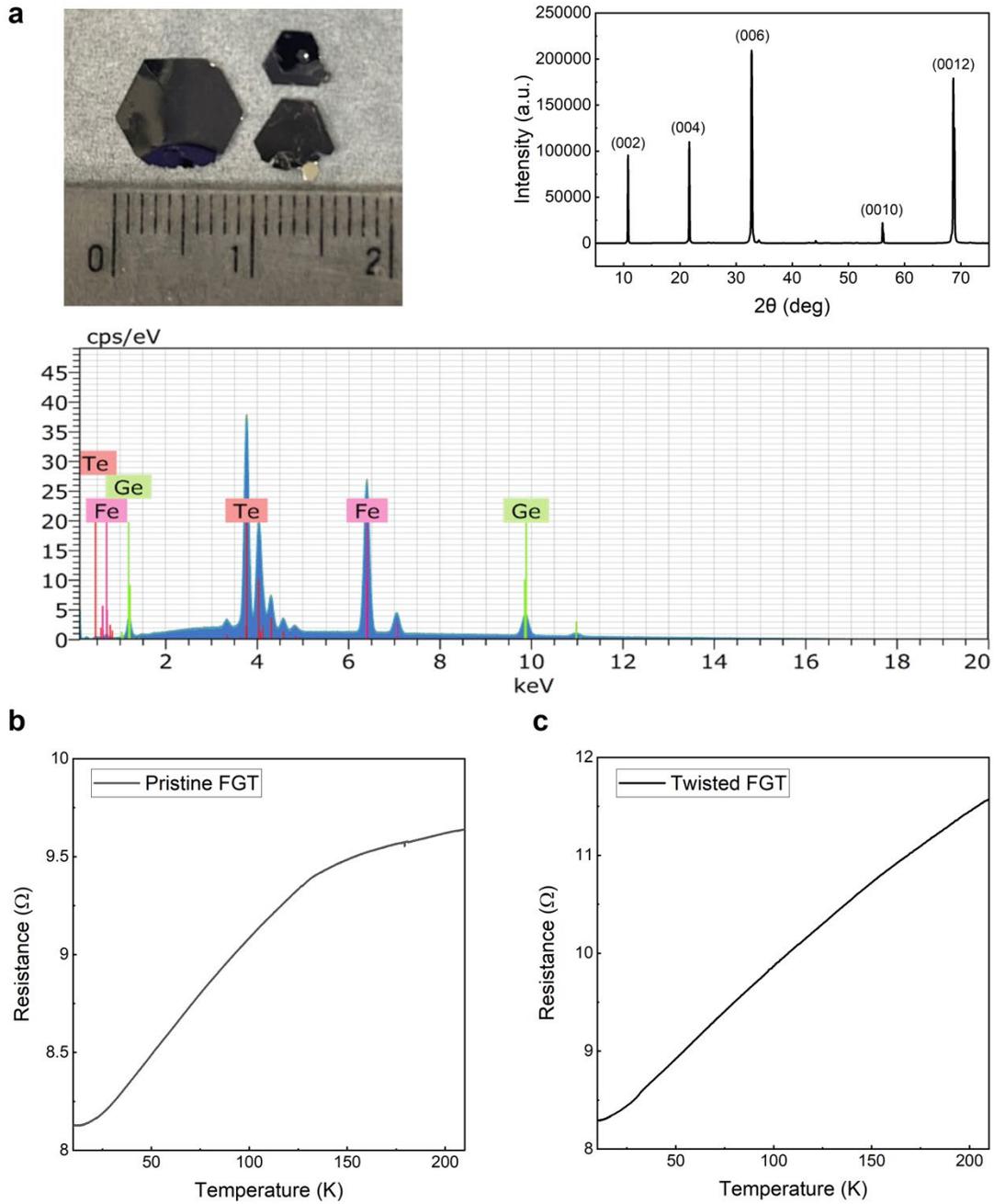

**Fig. S13 | Basic physical properties of the FGT samples. a** An optical image of the FGT single crystal, along with XRD and EDS data. **b, c** Temperature-dependent resistance of pristine and twisted FGT samples, respectively.

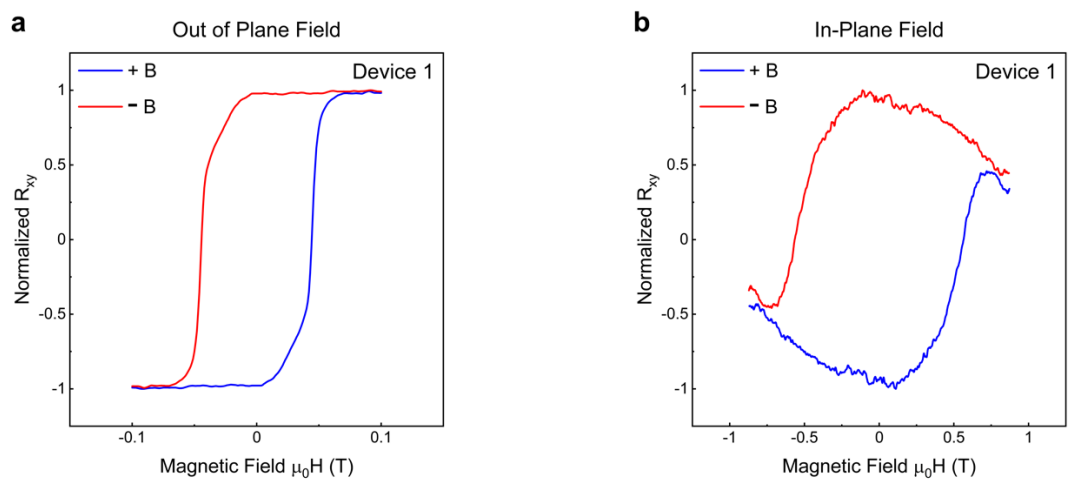

**Fig. S14 | Hysteresis loops of an isolated FGT nanoflakes. a** Hysteresis loop measured at 30 K with the magnetic field applied perpendicular to the plane. **b** Hysteresis loop measured at 30 K with the magnetic field applied parallel to the plane.

# Supplementary References